\documentstyle[graphicx,psfrag,eepic,11pt]{article}
\renewcommand{\baselinestretch}{1.5}
\newtheorem{theorem}{Theorem}[section]

\newtheorem{property}{Property}
\newtheorem{corollary}{Corollary}
\newtheorem{lemma}{Lemma}
 
\newtheorem{definition}{Definition}[section]

              {}

\def\squarebox#1{\hbox to #1{\hfill\vbox to #1{\vfill}}}
\newcommand{\qed}{\hspace*{\fill}
        \vbox{\hrule\hbox{\vrule\squarebox{.667em}\vrule}\hrule}\smallskip}

\newcommand{\comment}[1]{}

\newcommand{\ol}{\setlength{\itemsep}{0pt.}\begin{enumerate}}
\newcommand{\eol}{\end{enumerate}\setlength{\itemsep}{-\parsep}}
\def\squarebox#1{\hbox to #1{\hfill\vbox to #1{\vfill}}}

\newcommand{\lbl}[1]
  {\label{#1}}
\newcommand{\LE}{\hfill $\Box$} 

\newenvironment{code}[1]
    {\begin{list}%
        {{#1}.\arabic{LineNum}}%
        {\usecounter{LineNum}
                \setlength{\itemsep}{0in}
          }}%
    {\end{list}}

\newcommand{\thref}[1]{\mbox{Theorem~\ref{#1}}}

\newcommand{\lemref}[1]{\mbox{Lemma~\ref{#1}}}

\newcommand{\figref}[1]{\mbox{Figure~\ref{#1}}}
\newcommand{\eqref}[1]{\mbox{Equation~\ref{#1}}}

\newcounter{LineNum}

\newcommand{\I} {\item\hspace*{1em}}
\newcommand{\II} {\item\hspace*{2em}}
\newcommand{\III} {\item\hspace*{3em}}

\newcommand{\Rem}[1]   {$\ll$ {\it #1} $\gg$}

    {\begin{Sbox}\begin{minipage}}%
    {\end{minpage}\end{Sbox}\fbox{\TheSbox}}

\begin{document}
\pagenumbering{Roman}

\begin{titlepage}
\title{
Distributed Algorithms in Multihop Broadcast Networks}
\author{Israel Cidon \\ Department of Electrical Engineering \\
Technion - Israel Institute of Technology \\  cidon@tera.technion.ac.il
\and Osnat Mokryn \footnote{This work was done while the author was at the 
Department of Electrical Engineering, Technion.} 
\footnote{Correspondent author}
\\ 
Department of Computer Science \\ Hebrew University of Jerusalem \\
 osnaty@cs.huji.ac.il } 

\maketitle

\begin{abstract} 
The paper addresses the problem of solving classic distributed
algorithmic problems under the practical model of Broadcast 
Communication Networks. 
Our main result is a new Leader Election algorithm, with \mbox{$O(n)$} 
time complexity and \mbox{$O(n \cdot \lg(n))$} message 
transmission complexity. 
Our distributed solution uses a special form of the propagation 
of information with feedback (PIF) building block tuned to the 
broadcast media, 
and a special {\em counting and joining} approach for the 
election procedure phase. The latter is required for achieving 
the linear time. \\
It is demonstrated that the broadcast model requires
solutions which are different from the classic point to point model. 
\end{abstract}
\end{titlepage}
\section{Introduction}
\label{sec:intro} 
\vspace{-4mm}
Broadcast networks are often used in modern communication systems.
 A common broadcast network is a single hop shared media system where 
a transmitted message is heard by all nodes. Such networks include local 
area networks like Ethernet and token-ring, as well as satellite and 
radio networks.
In this paper we consider a more complex environment, in which a 
transmitted message is heard only by a group of neighboring nodes. 
Such environments include: Multihop packet radio networks, 
discussed for example in \cite{CS89}, \cite{CW91}; 
Multichannel networks, in which nodes may communicate via several 
non-interfering communication channels at different bands \cite{MaR83}; 
and a wireless multistation backbone system for
mobile communication \cite{BDS87}. \\
Since such networks are very important in the emerging area of 
backbone and wireless networks, it is important to design efficient 
algorithms for such environments.
We address here the problem of finding efficient algorithms for 
classic network problems such as 
propagation of information and leader election in the new models.

In the classic model of network communication, the problem of leader 
election is reducible to the problem of finding a spanning tree. 
The classic model is a graph of $n$ nodes and $m$ edges,
with the nodes representing computers that communicate via the 
edges which represent point-to-point bidirectional links.
Gallager, Humblet and Spira introduced in their pioneering 
work \cite{GHS83} a distributed minimum weight spanning tree (MST) 
algorithm, with \mbox{$O(n \cdot \lg(n))$} time and 
\mbox{$O(n \cdot \lg(n) + 2 \cdot m)$} message complexity. 
This algorithm is based on election phases in which the number 
of leadership candidates (each represents a fragment) is at least halved. 
Gallager et al. ensured a lower bound on a fragments level. 
In a later work, Chin and Ting \cite{CT85} improved Gallager's 
algorithm to \mbox{$O(n \cdot lg^*(n))$} time, estimating the 
fragment's size and updating its level accordingly, thus making a 
fragment's level dependent upon its estimated size. 
In \cite{Awe87}, Awerbuch proposed an optimal \mbox{$O(n)$} time and 
\mbox{$O(n \cdot lg(n) + 2 \cdot m)$} message complexity algorithm, 
constructed in three phases. In the first phase, the number of 
nodes in the graph is established. In the second phase, a MST is built 
according to Gallager's algorithm, until the fragments reach the 
size of \mbox{$n/\lg(n)$.} Finally, a second MST phase is performed, 
in which waiting fragments can upgrade their level, thus addressing a 
problem of long chains that existed in \cite{GHS83}, \cite{CT85}. 
A later article by Faloutsos and Molle (\cite{FM95}) addressed potential 
problems in Awerbuch's algorithm. In a recent work Garay, Kutten and Peleg
(\cite{GKP98}) suggested an algorithm for leader election in \mbox{$O(D)$} 
time, where $D$ is the diameter of the graph.
In order to achieve the $O(D)$ time, they use two phases. The first is
a controlled version of the GHS algorithm. 
The second phase uses a centralized algorithm, which concentrates on 
eliminating candidate edges in a pipelined approach. The message complexity
of the algorithm is \mbox{$O(m+ n \cdot \sqrt{n}$)}.

It is clear that all of the above election algorithms 
are based on the fact that sending different messages to
distinct neighbors is as costly a sending them the same message, 
which is not the case in our model. 
Our model enables us to take advantage of the broadcast topology, 
thus reducing the number of sent messages and increasing parallelism 
in the execution. Note, that while we can use distinct transmissions 
to neighbors it increases our message count due to unnecessary reception 
at all neighbors.  \\
Algorithms that are based on the GHS algorithm, chose a leader 
via constructing a MST in the graph. First these algorithms 
distinguish between internal and external fragment edges, 
and between MST-chosen and rejected edges.
An agreement on a minimal edge between adjacent fragments 
is done jointly by the two fragments, while other adjacent fragments
may wait until they join and increase in level.
In this paper it is seen that the broadcast environment requires
a different approach, that will increase parallelism in the graph.

Our main goal is to develop efficient distributed algorithms
for the new model. We approach this goal in steps. 
First, we present an algorithm for the basic task of Propagation of 
Information with Feedback (PIF) \cite{Seg83} with \mbox{$O(n)$} 
time and message transmission complexity. 
In the classic point-to-point model the PIF is an expensive building block
due to its message complexity. The native broadcast enables us to
devise a message efficient {\em fragment-PIF} algorithm, which
provides a fast communication between clusters.
Next, using the {\em fragment-PIF} as a building block, we present a 
new distributed algorithm for Leader Election, with  
\mbox{$O(n)$} time and \mbox{$O(n \cdot \lg(n))$} message 
transmission complexity.
In order to prove correctness and establish the time and message complexity, 
we define and use an equivalent high level algorithm for fragments, 
presented as a state machine. \\
The paper is constructed as follows: 
Section 2 defines the the model. Section 3
presents a PIF algorithm suited for the model. 
Section 4 introduces a distributed Leader 
Election algorithm for this model and shows and proves properties of 
the algorithm. Section 5 presents some simulation results of the 
distributed leader election algorithm. 
We conclude with a summary of open issues.
\vspace{-4mm}
\section{The Model} \lbl{sec:model} 
\vspace{-4mm}
A broadcast network can be viewed as a connected graph \mbox{$G(V,E)$,} 
where $V$ is the set of nodes. 
Nodes communicate by transmitting messages. 
If two nodes are able to hear each other's transmissions, 
we define this capability by connecting them with an edge. 
A transmitted message is heard only by a group of neighboring nodes. 
In this paper we use the terms message and message transmission 
interchangeably.
 $E$ is the set of edges. All edges are bidirectional. 
In the case of radio networks, we assume equal transmission 
capacity on both sides. \\
Our model assumes that every node knows the number of its neighbors. 
The originator of a received message is known either 
by the form of communication, or by indication in the message's header.\\
In the model, a transmitted message arrives in arbitrary final time 
to all the sender's neighbors. 
Consecutive transmissions of a node arrive to all its neighbors 
in the same order they were originated, and without errors. 
We further assume that there are no link or node failures, and additions. \\
It should be noted, that we assume that the media access and data link
problems, which are part of OSI layer 2 are already solved. 
The algorithms presented here are at higher layers, 
and therefore assume the presence of a reliable data link protocol which
delivers messages reliably and in order.
Bar-Yehuda, Goldreich and Itai (\cite{BGI92}) have addressed a lower level
model of a multihop radio environment even with no 
{\em collision detection} mechanism. In their model, concurrent 
receptions at a node are lost.
We assume models which are derived from conflict free allocation networks
such as TDMA, FDMA or CDMA cellular networks, which maintain a concurrent 
broadcast environment with no losses. 
\vspace{-4mm}
\section{Basic Propagation of Information Algorithms in our Model}
\label{sec:pif}
\vspace{-4mm}
The problem introduced here is of an arbitrary node that has a message it
wants to transmit to all the nodes in the graph. 
The solution for this problem for the classic model of 
communication networks was introduced by \cite{Seg83}, and is called 
Propagation of Information (PI). The initiating node is ensured 
that after it has sent the message to its neighbors, all the nodes 
in the network will receive the message in finite time. 
An important addition to the PI algorithm is to provide 
the initiator node with 
knowledge of the propagation termination, i.e., when it is ensured 
that all the nodes in the network have received the message. 
This is done with a feedback process, also described in \cite{Seg83} 
and added to the PI protocol. 
We describe a Propagation of Information with Feedback (PIF) 
algorithm for broadcast networks.
Because of the unique character of broadcast networks, it is 
very easy to develop a PI 
algorithm for this environment. When a node gets a message for 
the first time it simply 
sends it once to all neighbors, and then ignores any additional 
messages. In the feedback process 
messages are sent backwards over a virtual spanned tree in the 
broadcast network, to the initiator node.
\vspace{-4mm}
\subsection{Algorithm Description} \lbl{sec:Pdesc}
\vspace{-4mm}
We describe here the Propagation of Information with Feedback.
 A message in this algorithm is of the form: {\em MSG(target, l, parent)}, 
where {\em target} specifies the target node or nodes. 
A null value in the {\em target} header field indicates a
broadcast to all neighboring nodes, and is used when broadcasting 
the message. The {\em parent} field specifies the identity of the 
parent of the node that sends the message. 
It is important to note that a node receives a message only when 
addressed in the {\em target} header field by its 
{\em identification} number or when this field is null.
The $l$ field determines the sender's identity. The initiator, called the
{\em source} node, broadcasts a message, thus starting the propagation.  
Each node, upon receiving the message for the first time, 
stores the {\em identity} of the sender from which it got the message, 
which originated at the {\em source}, and broadcasts the message.
The feedback process starts at the {\em leaf} nodes, which are childless 
nodes on the virtual tree spanned by the PIF algorithm.
A {\em leaf} node that is participating in the propagation from 
the {\em source}, and has received 
the message from all of its neighboring nodes, sends back an 
acknowledgment message, called a feedback message, which is directed 
to its {\em parent} node. 
A node that got feedback messages from all of its child nodes, 
and has received the broadcasted message from all of its neighboring 
nodes sends the feedback message to its parent.
The algorithm terminates when the {\em source} node gets broadcast 
messages from all of its neighboring nodes, and feedback messages 
from all of its child neighboring nodes. \\
Formal description of the algorithm can be found in Appendix B.
\vspace{-4mm}
\subsection{Properties of the Algorithm}
\label{sec:Pprop}
\vspace{-4mm}
We define here the properties of the PIF algorithm in a broadcast network. 
Because of the similarity to the classic model, the time and message 
complexity is $O(n)$. 
\vspace{-2mm}
\begin{theorem} \lbl{th:Pprop}
Suppose a {\em source} node $i$ initiates a propagation of a message at 
time $t=\tau$. Then, we can say the following:
\begin{itemize}
\vspace{-5mm}
\item All nodes $j$ connected to $i$ will receive the message in finite time.
\vspace{-3mm}
\item Each node in the network sends one message during the propagation, 
and one message during the acknowledgment, to the total of two messages.
\vspace{-3mm}
\item The {\em source} node $i$ will get the last feedback 
message at no later than $\tau +2 \cdot n$ time units.
\vspace{-3mm}
\item The set of nodes formed by the set $\{parent_{s} \cup i\}$ 
nodes spans a virtual tree of fastest routes, from the {\em source} node, 
on the graph.
\end{itemize}
\end{theorem}
\vspace{-5mm}
The proof is similar to~\cite{Seg83}. 
\vspace{-5mm}
\section{Leader Election}
\lbl{sec:ldr}
\vspace{-4mm}
The leader election algorithm goal is to mark a single node in the graph 
as a leader and to provide its identity to all other nodes.
\vspace{-4mm}
\subsection{The Algorithm}
\lbl{sec:LdrAlg}
\vspace{-4mm}
During the operation of the algorithm the nodes are partitioned into 
{\em fragments}. 
Each fragment is a collection of nodes, consisting of a {\em candidate} 
node and its domain of supportive nodes. 
When the algorithm starts all the candidates in the graph are {\em active}. 
During the course of the algorithm, a candidate may become {\em inactive}, 
in which case its fragment joins an active candidate's fragment and no 
longer exists as an independent fragment. The algorithm 
terminates when there is only one candidate in the graph, 
and its domain includes all of the nodes. 
First, we present a higher level algorithm that operates at the fragment 
level. We term this algorithm the general algorithm. 
We then present the actual distributed algorithm by elaborating 
upon the specific action of individual nodes. 
In order to establish our complexity and time bounds we 
prove correctness of the general algorithm, and then prove that the 
general and distributed leader election algorithms are equivalent. 
We do so by proving that every execution of the 
general algorithm, specifically the distributed one, 
behaves in the same manner. 
We conclude by proving the properties and correctness of the algorithm.
\vspace{-6mm}
\subsubsection{The General Algorithm for a Fragment in the graph} 
\lbl{sec:Lgeneral}
\vspace{-4mm}
 We define for each fragment an {\em identity}, denoted by $id(F)$, 
and a {\em state}. 
The identity of a fragment consists of the size of the fragment, 
denoted by {\em id.size} and the candidate's identification number, 
denoted by {\em id(F).identity}. 
The state of the fragment is either {\em work}, {\em wait} or {\em leader}. 
We associate two variables with each edge in the graph, its current state 
and its current direction. An edge can be either in the state 
{\em internal}, in which case it connects two nodes that belong to the 
same fragment, or in the state {\em external}, when it connects two nodes 
that belong to different fragments. External edges will be directed in the 
following manner: Let $e$ be an external edge that connects two different 
fragments  $F1$ and $F2$. The algorithm follows 
these definitions for directing an edge in the graph: 
\vspace{-4mm}
\begin{definition} 
The lexicographical relation \mbox{\(id(F1)>id(F2)\)} holds if:  
\mbox{\(id(F1).size>id(F2).size\)} \\
 or if  \mbox{\([id(F1).size=id(F2).size\)} and 
\mbox{\(id(F1).identity>id(F2).identity]\)}.\lbl{df:edge1} 
\end{definition}
\vspace{-6mm}
\begin{definition}  
Let $e$ be the directed edge $(F1,F2)$ if \( id(F1)>id(F2)\)   
as defined by definition~\ref{df:edge1} \lbl{df:edge2} 
\end{definition}
\vspace{-5mm}
If the relation above holds,  $e$ is considered an {\em outgoing} edge 
for fragment $F1$ and an {\em incoming} edge for fragment $F2$, and 
fragments $F1$ and $F2$ are considered neighboring fragments. 
We assume that when an edge changes its direction, it does so in zero time.

\begin{figure}[htb]
\vspace{70mm}
\includegraphics{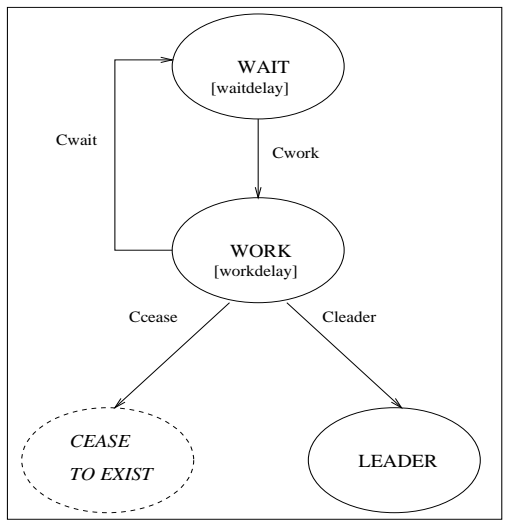}

\caption{\textit{A Fragment State Machine}} \lbl{fig:FSM}

\end{figure}

When the algorithm starts, each node is an active candidate, 
with a fragment size of 1. 
We describe the algorithm for every fragment in the graph by a state 
machine as shown in \figref{fig:FSM}. 
A fragment may be in one of the following states: {\em wait, work} or 
{\em leader}. A Fragment is in the virtual state {\em cease-to-exist} 
when it joins another candidate's fragment. The initial state for 
all fragments is {\em wait}, and the algorithm terminates when there is 
a fragment in the {\em leader} state. During the course of the algorithm, 
a fragment may move between states only when it satisfies the 
transition condition, as specified by the state machine.\\
The delays within the states are defined as follows:
 {\em waitdelay} - A delay a fragment suffers while in the {\em wait} 
state, while waiting for other fragments to inform of their {\em identity}. 
{\em workdelay}- This is a delay each fragment suffers while in the 
{\em work} state. 
Both delays are arbitrary limited, positive delays. 
The transition conditions are: {\em Cwait} which is defined by 
rule~\ref{en:Fbgr} below, {\em Cleader} which is defined by 
rule~\ref{en:Fleader} below, {\em Ccease} which is defined by 
rule~\ref{en:Fsmlr} below and {\em Cwork} which is defined by 
rule~\ref{en:Fwait} below. None of the transition conditions can 
cause any delay in time.
The transition conditions are the following:
 {\em Cwait}- The transition condition from the {\em work} state to the 
{\em wait} state. 
The condition is defined by rule~\ref{en:Fbgr} below. 
{\em Cleader}- The transition condition from {\em work} state to 
{\em leader} state. The condition is defined by rule~\ref{en:Fleader} below.
{\em Ccease}- The transition condition from the {\em work} state to 
the {\em cease-to-exist} virtual state. The condition is defined by 
rule~\ref{en:Fsmlr} below. {\em Cwork} - The transition condition 
from {\em wait} state to {\em work} state. 
The condition is defined by rule~\ref{en:Fwait} below. 
\vspace{-4mm}
\paragraph{The State Machine Formal Description:}
    \begin{enumerate}
    \vspace{-4mm}
    \item A fragment $F$ enters the {\em wait} state when it has at 
          least one {\em outgoing} edge ({\em Cwait} condition definition).
          \lbl{en:Fwait}
    \vspace{-4mm}
    \item A Fragment $F$ transfers to the {\em work} state from 
          the {\em wait} state ({\em Cwork} condition definition) 
          when all its external edges are {\em incoming} edges. 
          In the {\em work} state, the fragment will incur a delay 
          named {\em workdelay}, while it performs the 
          following: \lbl{en:Fwork}
      \begin{enumerate}
      \vspace{-4mm}
      \item Count the new number of nodes in its current domain. 
            The new size is kept in the variable {\em new\_size}. 
            We define the delay caused by the counting process by 
            {\em countdelay}.
            \lbl{en:Fnlvl}
    \vspace{-1mm}
      \item Compare its {\em new\_size} to the size of its maximal 
            neighbor fragment, $F'$.
            \footnote{Note, that before the action, $id(F') > id(F)$. 
              Therefore, $F'$ stays at its current size.}
        \begin{enumerate}
    \vspace{-1mm}
          \item If \mbox{$new\_size(F)> X \cdot id(F').size$} then
            fragment $F$ remains active. {\em ($X > 1$, a parameter.
            The optimal value of $X$ is calculated in 
            Section~\ref{sec:Lcomplx})}.\\ 
            Let $id(F).size \leftarrow new\_size$. $F$ changes 
            all of its external edges to the {\em outgoing} state. 
            (clearly, definition \ref{df:edge2} holds 
            here and at this step, all of its neighbors become aware of 
            its new size.) We define the delay caused by notifying the 
            neighboring fragments of its new size by {\em innerdelay}. 
            At this stage, condition {\em Cwait} is satisfied, 
            and $F$ transfers to the {\em wait} state. \lbl{en:Fbgr}
    \vspace{-1mm}
    \item Else, condition {\em Ccease} is satisfied, and $F$ ceases 
            being an active fragment 
            and becomes a part of its maximal neighbor fragment $F'$. 
            External edges between $F$ and $F'$ will become internal 
            edges of $F'$. $F'$ does not change its {\em size} or {\em id}, 
            but may have new external edges, which connect it through 
            the former fragment $F$ to other fragments. 
            The new external edges' state and direction are calculated 
            according to the current size of $F'$. 
            It is clear that all of them will be {\em outgoing} 
            edges at this stage. We define the delay caused by notifying 
            all of the fragment's nodes of the new candidate id 
            {\em innerdelay}. \lbl{en:Fsmlr}
        \end{enumerate}
      \end{enumerate}
\vspace{-2mm}     
    \item A Fragment that has no external edges is in the {\em leader} 
          state. ({\em Cleader} condition definition). \lbl{en:Fleader}
    \end{enumerate}
\vspace{-6mm}
\subsubsection{The Distributed Algorithm for Nodes in the Graph} 
\lbl{sec:Ldistribute}
\vspace{-4mm} 
We describe here the distributed algorithm for finding a leader in 
the graph. In Theorem~\ref{th:Lcnsst} we prove that this algorithm 
is equivalent to the general algorithm presented above.\\
When the algorithm starts, each node in the graph is an active 
{\em candidate}. During the course of the algorithm, the nodes are 
partitioned into fragments, supporting the fragment's {\em candidate}. 
Each node always belong to a certain fragment. Candidates may become 
{\em inactive} and instruct their fragment to join other fragments in 
support of another {\em candidate}. A fragment in the {\em work} state
may remain active if it is $X$ times bigger than its maximal neighbor.
The information within the fragment is 
transfered in PIF cycles, originating at the {\em candidate} node. 
The feedback process within each fragment starts at nodes called 
{\em edge} nodes. An {\em edge} node in a fragment 
is either a {\em leaf} node in the spanned PIF tree within the fragment, 
or has neighboring nodes that belong to other fragments. 
The algorithm terminates when there is one fragment that spans all the 
nodes in the graph.
\vspace{-2mm}
\begin{definition}
Let us define a PIF in a fragment, called a {\em fragment-PIF},
in the following manner: 
\begin{itemize}
\vspace{-4mm} 
\item The {\em source} node is usually the {\em candidate} node. 
  All nodes in the fragment recognize the candidate's identity. 
  When a fragment decides to join another fragment, the {\em source} node 
  is one of the {\em edge} nodes, which has neighbor nodes in the 
  joined fragment (the winning fragment).   
\vspace{-2mm} 
\item All the nodes that belong to the same fragment broadcast the 
  fragment-PIF message which originated at their {\em candidate's} node, 
  and no other fragment-PIF message.
\vspace{-2mm} 
\item An {\em edge} node, which has no child nodes in its fragment, 
  initiates a feedback message in the fragment-PIF when it has received 
  broadcast messages from all of its neighbors, either in its own fragment 
  or from neighboring fragments.
\vspace{-2mm} 
\end{itemize}
\lbl{df:LDpif}
\end{definition}
 
\textbf{Algorithm Description} \\
The algorithm begins with an initialization phase, in which every node,
which is a fragment of size $1$, broadcasts its identity.
During the course of the algorithm, a fragment that all of its {\em
edge} nodes have heard PIF messages from all their neighbors enter
state {\em work}. The fragment's nodes report back to the {\em
candidate} node the number of nodes in the fragment, and the
identity of the maximal neighbor. During the report, the nodes also
store a path to the edge node which is adjacent to the maximal
neighbor. The {\em candidate} node, at this stage also the {\em source} node,
compares the newly counted fragment size to
the maximal known neighbor fragment size. 
If the newly counted size is {\em not} at least $X$ times bigger than 
the size of the maximal neighbor fragment, then the
fragment becomes {\em inactive} and joins its maximal neighbor.
It is done in the following manner:
The {\em candidate} node sends a message, on the stored path, to the
fragment's edge node. This edge node becomes the fragment's {\em source}
node. It broadcasts the new fragment identity, which is the joined
fragment identity. At this stage, neighboring nodes of the joined
fragment disregard this message, thus the maximal neighbor fragment
will not enter state {\em work}. The {\em source} node chooses one of
its maximal fragment neighbor nodes as its parent node, and later on
will report to that node. From this point, the joining fragment
broadcasts the new identity to all other neighbors.
In case the newly counted size was at least $X$ times that of the
maximal neighboring fragment, the {\em candidate} updates the
fragment's identity accordingly, and broadcasts it. (Note, this is actually
the beginning of a new {\em fragment-PIF} cycle.)
The algorithm terminates when a {\em candidate} node learns that it
has no neighbors. \\
Appendix C contains a more detailed description, as well as a mapping
of the transition conditions and the delays in the general algorithm to 
the distributed algorithm.
\vspace{-2mm}
\subsection{Properties of the Algorithm}
\label{sec:Lprop}
\vspace{-2mm}
In this section we prove that a single candidate is elected in every 
execution of the algorithm. Throughout this section we refer to the 
high level algorithm, and then prove consistency between the versions.
All proofs of lemmas and theorems in this section appear in Appendix A.
\vspace{-3mm}
\begin{theorem} \lbl{th:Lcrct}
If a nonempty set of candidates start the leader election algorithm, 
then the algorithm eventually terminates and exactly one candidate is 
known as the leader.
\end{theorem}
\vspace{-6mm}
\begin{theorem} \lbl{th:Lcnsst}
If all the nodes in the graph are given different identities, 
then the sequence of events in the algorithm  does not depend on 
state transition delays.
\end{theorem}
\vspace{-6mm}
\begin{corollary} \lbl{co:Lcnsst1}
If all the nodes in the graph are given different identities, 
then the identity of the leader node will be uniquely defined 
by the high level algorithm.
\end{corollary}
\subsection{Time and Message Complexity} \lbl{sec:Lcomplx}
\vspace{-4mm}
We prove in this section, that the time complexity of the algorithm is
 $O(n)$. It is further shown, that for $X=3$ the time bound is the minimal,
and is $9 \cdot n$.
We also prove that the message complexity is $O(n \cdot lg(n))$.\\
In order to prove the time complexity, we omit an initialization phase, 
in which each node broadcasts one message. This stage is bounded by $n$ time units. 
(Note, that a specific node is delayed by no 
more than 2 time units - its message transmission, followed by its 
neighbors immediate response.)\\
Let us define a {\em winning} fragment at time $t$ during the execution of the 
algorithm to be a fragment which remained active after being in 
the {\em work} state. Thus, excluding the initialization phase, 
we can say the following:
\vspace{-4mm}
\begin{property} \lbl{pr:LDWDsize}
From the definitions and properties in Appendix C, it is clear that the delay 
{\em workdelay} for a fragment is bounded in time as follows: 
(a) For a fragment that remains {\em active}, the delay is the sum of 
{\em countdelay} and {\em innerdelay}. (b) For a fragment that becomes 
{\em inactive}, it is the sum:
\mbox{{\em countdelay} $+ 2 \cdot$ {\em innerdelay}. }
\end{property}
\vspace{-4mm}

\begin{theorem} \lbl{th:LXtime}
Let \mbox{$t^j_i$} be the time measured from the start, in which a 
{\em winning} fragment \mbox{$F_j$} completed the {\em work} state at 
size \mbox{$k_i$}. Then, assuming that 
\mbox{$workdelay_i^j \leq 3 \cdot k_i$,} and that 
\mbox{$countdelay^j_i \leq k_i$,}
we get that:
\vspace{-2mm}
\[t^j_i \leq \frac{X^2 + 3 \cdot X}{X-1} \cdot k_i\]
\vspace{-2mm}
\end{theorem}
\vspace{-11mm}
\paragraph{Proof of Theorem~\ref{th:LXtime}}
\vspace{-2mm}
We prove by induction on all times in which fragments in the graph 
enter state {\em work}, denoted \mbox{$\{t^{j}_{i}\}$.} \\
\mbox{$t_1$:} The minimum over \mbox{$t_i^j$} for all $j$.
It is clear that the lemma holds.\\
Let us assume correctness for time \mbox{$t_i^j$,} 
and prove for time \mbox{$t_i^j+workdelay_i^j$,} for any $j$. \\
From lemma~\ref{le:Lsze} we deduct that \mbox{$k_i > X^2 \cdot k_{i-1}$.} 
Therefore, according to the induction assumption, \mbox{$F_j$} entered 
state $work$ for the $i-1$ time in time:  
$t^j_{i-1} \leq \frac{X^2 + 3 \cdot X}{X-1} \cdot k_{i-1}$.
Let us now examine the time it took \mbox{$F_j$} to grow from its size 
at time \mbox{$t^j_{i-1}$} to size \mbox{$k_i$,}: \mbox{$t^j_i-t^j_{i-1}$.} 
Since \mbox{$F_j$} is a winning fragment at time \mbox{$t^j_i$,} 
it is at least $X$ times the size of any of its neighboring fragments. 
In the worst case, the fragment has to wait for a total of
size \mbox{$(X-1) \cdot \frac{k_{i}}{X}$} that joins it.
Note, that the induction assumption holds for all of $F_j$'s neighbors.
So we stated that: 
\begin{enumerate}
\vspace{-4mm}
\item The time it took a winning fragment to get to size \mbox{$k_{i-1}$} is 
  \mbox{$t^{j}_{i-1}$}, and  \mbox{$k_{i-1} \leq 1/X^2 \cdot k_{i}$}
  \lbl{en:LptimeX1}
  \vspace{-2mm}
\item The fragment then entered the work state to discover that its
  actual size is \mbox{$k^{`j}_{i-1} \leq 1/X \cdot k_{i}$} 
  \vspace{-2mm}
\item $F$ has to wait for joining fragments, in the overall size of: 
  \mbox{$(X-1) \cdot 1/X \cdot k_{i}$}
\vspace{-2mm}
\item The time it takes the fragment's candidate of size $k_{i}$ to be 
  aware of its neighboring fragments identities and its own size while in 
  the {\em work} state is bounded by 
  \mbox{$countdelay_i^j \leq k_{i}$.} 
  \lbl{en:LptimeX4}
 \end{enumerate}
\vspace{-4mm}
From items~\ref{en:LptimeX1} to~\ref{en:LptimeX4} we can conclude that:\\ 
\begin{math} 
t^{j}_{i} \leq \frac{X^2 + 3 \cdot X}{X-1} \cdot k_{i-1} + workdelay^j_{i-1} + 
\frac{X^2 + 3 \cdot X}{X-1} \cdot (X-1) \cdot \frac{1}{X} \cdot k_{i} + 
countdelay_i^j \\
~~~~\leq  \frac{X^2 + 3 \cdot X}{X-1} \cdot \frac{1}{X^2} \cdot k_{i} + 
3 \cdot \frac{1}{X} \cdot k_{i} + 
\frac{X^2 + 3 \cdot X}{X-1} \cdot (X-1) \cdot \frac{1}{X} \cdot k_{i} + k_{i}\\
~~~~=\frac{X^2 + 3 \cdot X}{X-1} \cdot k_{i} 
\end{math} 
\LE
\vspace{-3mm}
\begin{corollary}
From property~\ref{pr:LDWDsize} it is obtained that the delay 
{\em workdelay} for every fragment of size $k$ is indeed bounded 
by $3 \cdot k$, and that the delay {\em countdelay} is
bounded by $k$. Therefore, from Theorem~\ref{th:LXtime} the time it 
takes the distributed algorithm to finish is also bounded by
\[t^j_i \leq \frac{X^2 + 3 \cdot X}{X-1} \cdot n \]
where $n$ is the number of nodes in the graph, and therefore is $O(n)$. 
\end{corollary}
\vspace{-7mm}
\begin{corollary} \lbl{co:LXbest}
When $X = 3$ the algorithm time bound is at its 
minimum: $t^j_i \leq 9 \cdot n$.
\end{corollary}
\vspace{-2mm}
\begin{figure}[bht]
\vspace{55mm}
\includegraphics{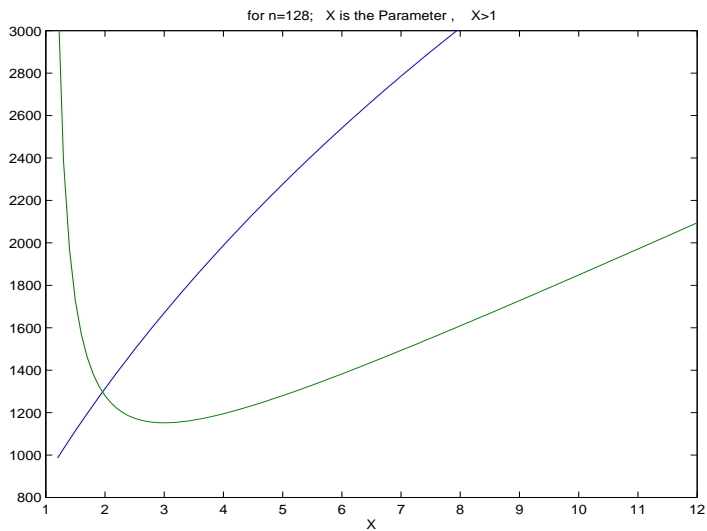}
\vspace{12mm}
\caption{\textit{Time and Message Complexity as a factor of X}} \lbl{fig:Ltmgc}
\end{figure}
\vspace{-5mm}
\begin{theorem} \lbl{th:LXmsg}
The number of messages sent during the execution of the algorithm is 
bounded by: 
\[( \frac{lg(n) - lg(1+X)}{lg(\frac{X+1}{X})} + 1) \cdot n
\hspace{2mm} ; \hspace{5mm} X>1 \].
\vspace{-4mm}
\end{theorem}
\vspace{-9mm}
The proof of Theorem~\ref{th:LXmsg} is in Appendix A.
\section{Simulation results}
\vspace{-4mm}
The Distributed Leader Election was simulated using SES/Workbench, which 
is a graphical event driven simulation tool. The algorithm was simulated for 
a multihop broadcast environment.
The algorithm was tested for a wide range of different topologies, among 
them binary trees, rings and strings, as well as more connected 
topologies. 
The simulation enables the user to choose the number of nodes, a basic
topology, the connectivity and the growth factor $X$. The simulation output
consists of the identity of the chosen leader node,
the total time it took to get to the decision (excluding initialization phase)
and the number of transmissions sent during the course of the algorithm. 

Our simulation program first generates a random topology for the operation 
of the algorithm. The method for building the topology is based on an 
initial simple connected graph (tree, string, ring, etc.) and a random 
addition of edges to that initial graph.
We present here results of two initial graphs: strings and binary trees.
The numbered nodes are randomly arranged into the given initial graph.
Then, adhering to the connectivity parameter and correctness, edges are 
randomly added to the graph. The connectivity parameter, $C$, is defined
as the ratio between the number of additional edges in the graph out of 
all possible edges. For each topology and connectivity 100 different 
graphs were generated and executed. 

We present here the maximal time and transmission
results measured for each topology and connectivity, for a fixed number 
of nodes and the same growth factor. 
\figref{fig:MsnkT} and \figref{fig:MbtrT} present the maximal measured
time for string and binary tree based topologies, respectively.
The slowest measured topology is a pure string, since it does not
exploit the broadcast ability. Minimal time for both topologies 
is measured for connectivity of 20\%-30\%. In both cases,
the measured time is less than the number of nodes, due to maximal
parallelism in the algorithm.
As connectivity grows, the algorithm loses parallelism.

\begin{figure}[hbt]
\vspace{10mm}
\vspace{50mm}
\includegraphics{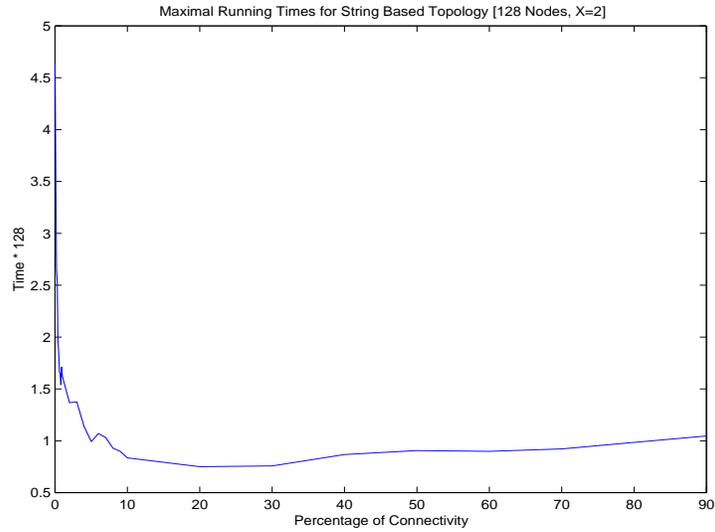}
\vspace{12mm} 
\caption{\textit{Maximal Times for String Based Topologies}}
\lbl{fig:MsnkT}
\end{figure}

\begin{figure}[hbt]
\vspace{55mm}
\includegraphics{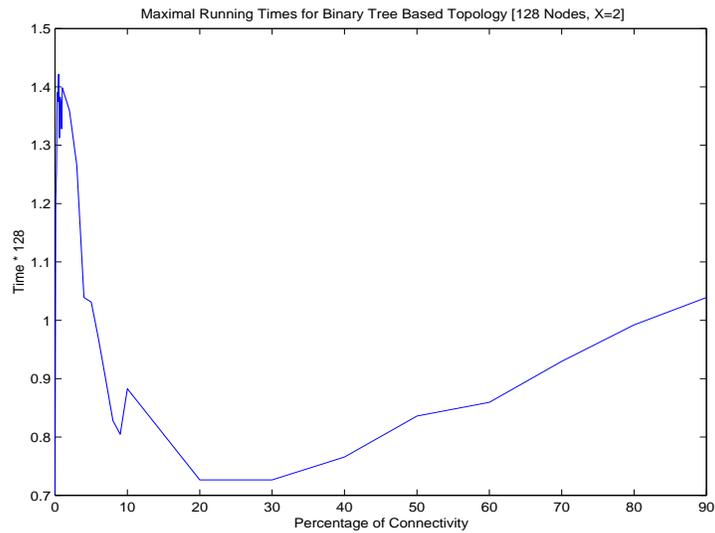}
\vspace{12mm} 
\caption{\textit{Maximal Times for Binary Tree Based Topologies}} 
\lbl{fig:MbtrT}
\vspace{2mm}
\end{figure}

Our results show that for connectivity of 20 \% and up,
the maximal identity node was chosen the leader of the graph in most cases, 
and from connectivity of 30\% and up it was always selected.
This scenario best suits the case of a fully connected graph, in 
which all nodes are aware of all possible identities 
as soon as the initialization phase is over. 
Thereafter, each node, starting with the minimal identity node, joins the
maximal identity node by order of increasing identities.
In the case of a fully connected graph, there is a strict increasing order of 
joining among the nodes. When connectivity is 20\% to 30\% , the
maximal identity node is well known in the graph, but parallelism is high.
This leads to a situation in which nodes join the maximal identity node
without waiting for each other, and the leader is chosen very quickly.\\\\


\begin{figure}[hbt]
\vspace{50mm}
\includegraphics{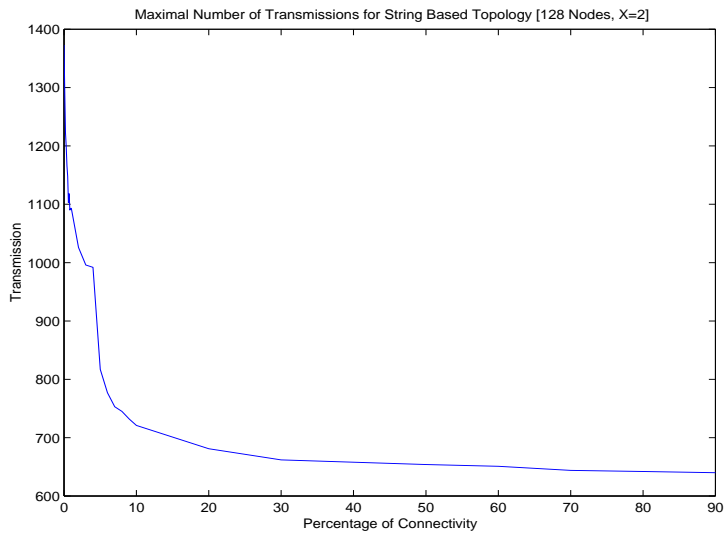}
\vspace{15mm} 
\caption{\textit{Maximal Transmissions for String Based Topologies}} 
\lbl{fig:MsnkM}
\end{figure}
\begin{figure}[htb]
\vspace{50mm}
\includegraphics{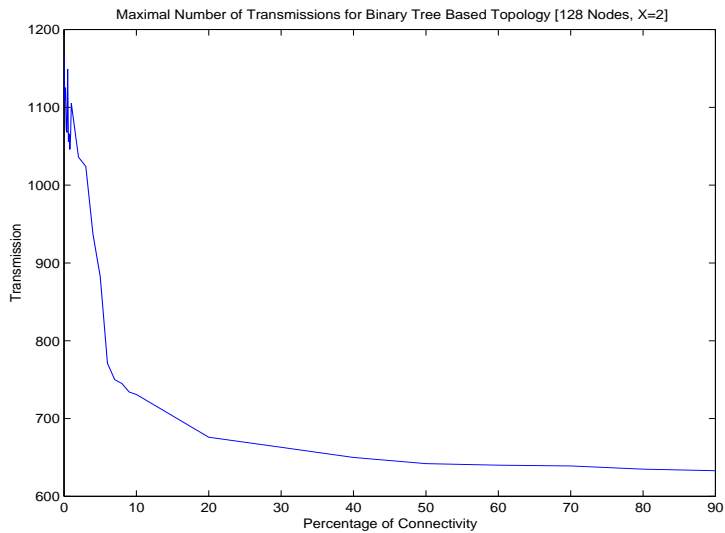}
\vspace{15mm} 
\caption{\textit{Maximal Transmissions for Binary Tree Based Topologies}} 
\lbl{fig:MbtrM}
\end{figure}

\figref{fig:MsnkM} and \figref{fig:MbtrM} present maximal measured 
transmissions for a string based topology and a binary tree based topology.
It can be observed, that generally higher connectivity means less 
transmissions.
This follows the order scheme discussed earlier for densed graphs,
in which the maximal identity node is known throughout 
the graph and rarely has any competition. This leads to a situation in which
nodes join the winning candidate very early in the algorithm,
and do not participate in unnecessary {\em work} states.

\section{Summary}
\vspace{-4mm}   
We presented new distributed algorithms for emerging modern 
broadcast communication systems.
Under the new model we introduced algorithms for PIF and Leader Election.
The algorithms are optimal in time and message complexity. \\
By using the {\em fragment-PIF} approach our leader election 
algorithm enables a fragment to affect all of its neighbors
concurrently, thus increasing parallelism in the graph.
This new approach may be used for all fragment-based decision algorithms
in a broadcast environment. 

There are several more problems to investigate under this model, 
such as other basic algorithms (e.g. DFS), failure detection and recovery.
Other environments, such as a multicast environment, may require different
approaches. 
Extensions of the model might also be viewed. Broadcast LANs are often 
connected via bridges. This leads to a more general model that includes 
both point to point edges and a broadcast group of edges connected to 
each node. For such a general model, the design of specific algorithms 
is required as well. 
\label{sec:sum}
\vspace{-4mm}
\section*{Acknowledgment}
\vspace{-2mm}
We would like to thank Dr. Avi Mendelson, Prof. Shmuel Zaks and Dr. Shay
Kutten for many helpful discussions and comments. \\
We would also like to thank Erez Brickner and Iris Shraibman for 
developing the simulation models and producing the simulation results.

\vspace{-2mm}
\renewcommand{\baselinestretch}{1} 
\normalsize
\bibliographystyle{alpha}
\bibliography{letr}

\clearpage
\renewcommand{\baselinestretch}{1.2} 
\appendix
\noindent 
\begin{center}
\textbf{\LARGE Appendix A} \\
\lbl{app:app:proofs}
\end{center}
\paragraph{Proof of \thref{th:Lcrct}} 
In order to proof the Theorem, we will state some Properties and
Lemmas.
\vspace{-4mm}
\begin{property} \lbl{pr:LEdgesOut}
When an active fragment changes its state from {\em work} to {\em wait}, 
all of its external edges are 
outgoing edges.
\end{property}
\vspace{-6mm}
\begin{lemma} \lbl{le:Lactv}
During every execution of the algorithm, at least one active candidate 
exists.
\end{lemma}
\vspace{-8mm}
\paragraph{Proof of \lemref{le:Lactv}}
By definition, an active candidate exists in the graph either in the
{\em work}, {\em wait} 
or {\em leader} state. A candidate becomes inactive according to the
algorithm if it 
discovers, while in the {\em work} state, an active neighbor fragment
of bigger size, as can be seen in~\ref{en:Fsmlr}. 
By this definition it is clear that a candidate may cease to exist
only if it encountered another 
active candidate of at least half its size.
Therefore, during the execution of the algorithm, 
there will always be at least one active candidate. \LE
\vspace{-4mm}
\begin{property} \lbl{pr:LEdgesOut}
When an active fragment changes its state from {\em work} to {\em wait}, 
all of its external edges are 
outgoing edges.
\end{property}
\vspace{-4mm}
\begin{lemma} \lbl{le:Lwork}
During the execution of the algorithm, there is always a fragment 
in the {\em work} or {\em leader} states.
\end{lemma}
\vspace{-8mm}
\paragraph{Proof of \lemref{le:Lwork}} 
A fragment will be in the {\em work} state if it has the minimal $id$
among all its neighbors. 
Definition~\ref{df:edge1} determines that the fragment's $id$ consists
of its size and identity 
number. If there exists a fragment of minimal size, then it is the
fragment with the 
minimal $id$. Otherwise, there are at least two fragments in the graph
that have the same 
minimal size. As defined, the fragment with the lower identity number
has the lower $id$. 
It is clear then, that at any stage of the algorithm there exists a
fragment with a minimal $id$, and therefore there will always be a
fragment in the {\em work} or 
{\em leader} state. \LE
\vspace{-4mm}
\begin{lemma} \lbl{le:Lordr}
If a fragment $F$ was in the {\em work} state and remained active 
in {\em wait} state, it will enter the {\em work} state again only 
after each and every one of its neighboring 
fragments has been in the {\em work} state as well.  
\end{lemma} 
\vspace{-8mm}
\paragraph{Proof of \lemref{le:Lordr}}
According to Property~\ref{pr:LEdgesOut}, when $F$ entered the
{\em wait} state, all of its edges are in the {\em outgoing} state. 
In order to enter the {\em work} state again, all of $F$'s {\em external} 
edges have to be {\em incoming} edges. While in the {\em wait} state, $F$ 
does nothing. Its neighbor activity is the only cause for a change in the 
edges' state or direction. An {\em outgoing} edge may change its state to 
{\em internal}, if the neighbor fragment enters the {\em work} state and 
becomes a supporter of $F$ or changes its direction to {\em incoming}, 
according to Definition~\ref{df:edge2}, 
as a result of a neighbor fragment changing its {\em size}. 
Hence, all $F$'s neighboring fragments must be in the {\em work} state 
before $F$ may enter it again.   
 \LE
\vspace{-4mm}
\begin{corollary} \lbl{co:Lordr}
A fragment $F$ in the {\em wait} state which has a neighbor 
fragment $F'$ in the {\em work} state may enter the {\em work} state 
only after $F'$ has changed its state to the {\em wait} state.
\end{corollary}
\vspace{-6mm}
\begin{lemma} \lbl{le:Lsze}
Let us consider a fragment, $F$, that entered the {\em work} state for 
two consecutive times, $i$ and \mbox{$i$+1} and remained active.
Let $t_{i}$ be the time it entered the {\em work} state for 
the $i$-$th$ time, and let \mbox{$t_{i+1}$} be the \mbox{$i$+1} time. 
Let us define by \mbox{$k_i$} its known size at time \mbox{$t_i$}, and by 
size \mbox{$k_{i+1}$} its size at time \mbox{$t_{i+1}$+ countdelay.} 
Then, \mbox{$k_{i+1} \geq X^2 \cdot k_{i}$.}
\end{lemma} 
\vspace{-8mm}
\paragraph{Proof of \lemref{le:Lsze}}
If $F$ enters the {\em work} state for the \mbox{$i$-$th$} time at
time \mbox{$t_{i}$,} then according to 
Lemma~\ref{le:Lordr} all of its neighbor fragments will be in the {\em work} 
state before time \mbox{$t_{i+1}$.} We know that at time \mbox{$t_{i}$} $F$ is
of known size \mbox{$k_i$.} 
According to rules~\ref{en:Fsmlr} and \ref{en:Fbgr}, if it stayed
active, it has enlarged its size by a factor of $X$ at least. (Note, $X>1$). 
Since $F$ reentered the {\em work} state, any of its new neighboring fragments 
is at least as big as F, i.e.~at size of at least \mbox{$X \cdot k_{i}$.} 
We know that $F$ remains active after completing both {\em work} states. 
Therefore, following the same rules [~\ref{en:Fsmlr} and \ref{en:Fbgr}~], at 
time \mbox{$t_{i+1}$+countdelay,} its size, \mbox{$k_{i+1}$,} must be
at least $X$ times the size of its maximal neighbor.
It follows directly that \mbox{$k_{i+1} \geq X^2 \cdot k_{i}$.} 
 \LE
\vspace{-4mm}
\begin{corollary} \lbl{co:Llvl}
The maximal number of periods that an active candidate will be in 
the {\em work} state is $\lg_X{n}$. 
\end{corollary}
\vspace{-4mm}
Now we can proceed with the Theorem: \\
From Lemma~\ref{le:Lactv} and Lemma~\ref{le:Lwork} it is clear that the 
algorithm does not deadlock, and that the set of active candidates is
non-empty throughout the course of the algorithm.
Property~\ref{co:Llvl} limits the number of times a candidate 
may enter the {\em work} state.
Since at any time there is a candidate in the {\em work} state, 
there always exists a candidate that enlarges its domain.
It follows that there exists a stage in the algorithm where there will
be only one candidate, and its domain will include all 
the nodes in the graph.  \LE
\vspace{-4mm}
\paragraph{Proof of \thref{th:Lcnsst}}
Let $U$ be the set of fragments in the {\em work} state at any arbitrary 
time during the execution of the general algorithm. 
If we view the times before a fragment in the graph enters 
the {\em leader} state, then by lemma~\ref{le:Lwork} and 
corollary~\ref{co:Lordr} we can conclude that:
\begin{enumerate}
\vspace{-4mm}
\item $U$ is non-empty until a leader is elected \lbl{ls:LUnonE} 
\vspace{-2mm}
\item Fragments in $U$ cannot be neighbors \lbl{ls:LUnonN}
\vspace{-2mm}
\item All the neighboring fragments of a fragment \mbox{$F \subseteq U$} 
  cannot change their {\em size} until $F$ is no longer in $U$ 
  (i.e., $F$ completes the {\em work} state and changes its state 
  to {\em wait}). \lbl{ls:LUNwait}
\end{enumerate}
\vspace{-4mm}
At every moment during the execution of the algorithm,
the graph is acyclic and the edges' directions determine uniquely
which fragments are in the {\em work} state. 
Let us examine the state of the graph at an arbitrary time,
before any fragment enters the {\em leader} state.
The conclusions above imply that there is a predetermined order in which 
fragments enter the {\em work} state. The time it takes a fragment to
enter the {\em work} state from the {\em wait} state is determined by
the {\em Cwork} condition. 
Excluding initialization, a fragment can reenter the {\em work} state
after each of its neighboring fragments completed a period in 
the {\em work} state. The duration within the {\em work} state for a
fragment, depends only on the local delay {\em workdelay} as defined
by~\ref{pr:LDWDsize}. Therefore, fragments that enter 
the {\em work} state complete it in finite time.
Therefore, it is clear, that the order by which the fragment's
neighbors enter the {\em work} state does not affect the order by which it 
enters the {\em work} state. This proves that the sequence of events
between every pair of neighboring fragments in the graph is
predetermined, given the nodes have different identities. 
Therefore, the sequence of events between all the fragments in the
graph is predetermined and cannot be changed throughout the execution
of the algorithm.
\LE
\vspace{-5mm}
\paragraph{Proof of Theorem~\ref{th:LXmsg}}
In order to prove the Theorem, we use the following Lemmas and properties.
\vspace{-3mm}
\begin{property}
During the execution of the algorithm, nodes within a fragment send 
messages only while in the {\em work} state. \lbl{pr:Lmsg}
\end{property}
\vspace{-8mm}
\begin{lemma} \lbl{le:LDWmsg}
While in the {\em work} state, the number of messages sent in a fragment 
is at most \mbox{$3 \cdot k$,} where $k$ is the number of nodes within 
the fragment.
\end{lemma}
\vspace{-10mm}
\paragraph{Proof of \lemref{le:LDWmsg}}
While in the {\em work} state, each node can send either one of three 
messages: FEEDBACK, ACTION and INFO (in this order). 
Every message is sent only once, from the PIF properties. Therefore, 
the number of messages sent in a fragment of size $k$ is bounded 
by \mbox{$3 \cdot k$.}
\LE
\vspace{-6mm}
\begin{lemma} \lbl{le:LXtmsg}
Let $l$ be the number of times a cluster of nodes, initially a fragment,
entered state {\em work}. Then: 
\vspace{-2mm}
\[ l = \frac{lg(n) - lg(1+X)}{lg(\frac{X+1}{X})} + 1\]
\end{lemma}
\vspace{-6mm}

\paragraph{Proof of \lemref{le:LXtmsg}}
A fragment of an arbitrary size $k$, may join a fragment of 
size $\frac{1}{X} \cdot k + 1$. Therefore, $ \forall l$, the
minimal growth rate is of the form: 
\(k_l = k_{l-1} + \frac{1}{X} \cdot k_{l-1}+1 = \frac{X+1}{X} \cdot k_{l-1}+1\)
, where $k_0 = 0$. \\
Let us define the following: 
\vspace{-4mm}
\[k_l = a \cdot k_{l-1} +1 ;\hspace{5mm} a= \frac{X+1}{X}; \hspace{5mm}
\sum_{l=1}^\infty k_l \cdot Z^l = k(Z)\]
\vspace{-2mm}
By using the $Z$ transform as defined above, it follows that: \\
\vspace{-2mm}
\[k_l = a \cdot \sum_{l=1}^\infty k_{l-1} \cdot Z^l = 
a \cdot Z \cdot \sum_{l=1}^\infty k_{l-1} \cdot Z^{l-1} = 
a \cdot Z \cdot k(Z)\]
\vspace{-2mm}
Since $k_0 = 0$ and \(\forall Z < 1, \sum_{l=0}^\infty Z^l = \frac{1}{1-Z}\),
we obtain that: 
\[k(Z) = a \cdot Z \cdot k(Z) + \frac{1}{1-Z} \hspace{3mm} \Rightarrow
\hspace{3mm} k(Z) = \frac{1}{(1-Z) \cdot (1-a \cdot Z)} \]
\vspace{-4mm}
By separation of variables we get:
\vspace{-2mm}
\[k(Z) = \frac{X+1}{1-a \cdot Z} - \frac{X}{1-Z}\]
\vspace{-4mm}
Therefore: 
\vspace{-2mm}
\[\sum_{l=1}^\infty k_l \cdot Z^l = (X+1) \cdot \sum_{l=1}^\infty 
(a \cdot l)^l - X \cdot \sum_{l=1}^\infty Z^l = 
\sum_{l=1}^\infty [(X+1) \cdot a^l -X] \cdot Z^l \]
Since by definition $a>1$, we get: \(k_l = (X+1) \cdot a^l - X \geq
a^l + X \cdot a^l -X \geq a^{l-1} + X \cdot a^{l-1} = (1+X) \cdot a^{l-1}.\)
We require that: \((1+X) \cdot a^{l-1} = n\), where $n$ is the number of nodes
in the graph. By taking the logarithm of both sides, we obtain:
\vspace{-2mm}
\[ l = \frac{lg(n) - lg(1+X)}{lg(\frac{X+1}{X})} + 1 \]
\vspace{-4mm}
\LE \\
Now we may proceed to prove the message complexity: \\
From Property~\ref{pr:Lmsg} it is clear that every fragment sends 
messages only while it is in the {\em work} state. Lemma~\ref{le:LDWmsg} 
shows that in the {\em work} state, the maximal number of messages sent 
within a fragment of size $k$ is \mbox{$3 \cdot k$.} From \lemref{le:LXtmsg}
and the above we obtain that the message complexity is:
\vspace{-2mm}
\[( \frac{lg(n) - lg(1+X)}{lg(\frac{X+1}{X})} + 1) \cdot n
\hspace{2mm} ; \hspace{5mm} X>1 \].
\LE

\clearpage
\appendix
\noindent 
\begin{center}
\textbf{\LARGE Appendix B} \\
\lbl{app:app:pifalg}
\textbf{Propagation of Information with Feedback Formal Description}
\end{center}
\vspace{-8mm}
\paragraph{Protocol messages}
\begin{description}
\vspace{-4mm}
\item [$START$] -- Upon receiving a $START$ message a node will initialize 
  its internal protocol variables and initiates propagation of an information 
  in the network. A $START$ message may be delivered to an undetermined number
  of nodes, each of which will initiates an independent propagation of 
  information in the network.
\vspace{-2mm}
\item [$MSG$]-- The message propagated in the network. The propagation comes 
  to an end when The starting nodes gets it back from all of its descendants 
  nodes.
\end{description}
\vspace{-2mm}
Each protocol message includes the following data:
\begin{description}
\vspace{-4mm}
\item [$target$] -- Specifies the target node or nodes. zero will indicate 
  all neighboring nodes.
\vspace{-2mm}
\item [$l$] -- The sender's identification.
\vspace{-2mm}
\item [$parent$] -- The sender's parent node. zero if none.
\end{description}
\vspace{-2mm}
Each Node uses the following variables:
\begin{description}
\vspace{-4mm}
\item[$i$] -- The node's identity
\vspace{-2mm}
\item[$l$] -- The message sender's identity
\vspace{-2mm}
\item[$parent$] -- The node's parent's identity (zero for a start node)     
\end{description}
\vspace{-4mm}
\paragraph{The algorithm for a starting node $s$}
\vspace{-4mm}
\begin{code}{s}
\vspace{-4mm}
\I For a $START(s,s,0)$ message:
\III set $m \leftarrow$ 1; set $parent \leftarrow$ 0; Send $MSG(0,s,parent)$
\I For a $MSG(target, \ell, parent)$ message:\\
\Rem {Message is ignored if $parent$=$s$ AND $target \neq s$}
\II if $parent$ = $s$ AND $target \neq s$ IGNORE 
\II else:
\III set $N(l) \leftarrow$ $1$ 
\III if  $\forall \ell~ N(\ell)=1$ END
\end{code}
\vspace{-4mm}
\paragraph{The Algorithm for a node i} 
\vspace{-4mm}
\begin{code}{i}
\vspace{-4mm}
\I For a $MSG(target, \ell, parent)$ message received at node i:\\
\Rem {Message is ignored if $parent$=$s$ AND $target \neq s$}
\II  if $parent$ = $s$ AND $target \neq s$ IGNORE 
\II else:
\III set $N(l) \leftarrow$ $1$ 
\III if $m$ = 0 : set $m \leftarrow$ 1; set $parent \leftarrow$ $\ell$; 
send $MSG(0, i, parent)$
\III if  $\forall \ell~ N(\ell)=1$ send $MSG(parent, i, parent)$ on the link 
to $parent$.  
\end{code}

\clearpage
\appendix
\noindent 
\begin{center}
\textbf{\LARGE Appendix C} \\
\lbl{app:app:LDRdst}
\textbf{Distributed Algorithm for Leader Election - Detailed
Description}
\end{center}
\vspace{-4mm}
\paragraph{The messages sent}
\vspace{-4mm}
[All messages include additional fragment-PIF header fields.]
\begin{enumerate}
\vspace{-2mm}
\item A broadcast INFO message from the {\em source} node of the fragment 
with the following data fields: A fragment's current {\em size} and 
{\em identity} and a field that denotes the former fragment's identity, 
for recognition within the fragment.  

\item A FEEDBACK message, sent during the feedback phase. 
The message data field includes the accumulated number of nodes under 
this node in the fragment. It also includes a flag, indicating 
whether another neighboring fragment was encountered. 
If this flag is set, then the next two fields indicate the maximal 
neighbor fragment's {\em size} and {\em identity}.

\item  An ACTION message, sent only in case the fragment's candidate 
observes that it has a neighboring fragment it should join. 
In this case it sends this message to one of 
its {\em edge} nodes, that has neighboring nodes in the joined fragment.
\end{enumerate}
\vspace{-4mm}
\paragraph{A detailed description} 
\begin{description}
\vspace{-2mm}
\item {\em Initialization:} The algorithm starts when an arbitrary number 
of nodes initialize their state and send an initialization INFO message, 
with their {\em identity}. Upon receiving 
such a message, a node sends its INFO message if it hasn't done so yet, 
and notes the maximal and minimal identity it has encountered. 
When a node has received such INFO messages from all of its neighbors, 
it determines its state. If the node has the minimal identity amongst all 
of its neighbors, then it is in the {\em work} state, 
otherwise it is in the {\em wait} state. 
A node that enters the {\em work} state joins its maximal neighbor, 
initializes all of its inner variables accordingly, and broadcasts its 
INFO message, stating its maximal neighbor as its parent. 
A node in the {\em wait} state initializes all of its inner variables.

\item {\em Propagation of INFO in a fragment:} 
All the nodes that belong to the candidate's fragment, 
upon first receiving a message that includes their fragment's 
{\em identity}, do the following: note the fragment's new {\em identity} 
as delivered in the message, record the identity of the node from which 
the message was first received in a {\em parent} variable, 
and broadcast the message. 

\item {\em Encountering another fragment:} An {\em edge} node, which 
receives for the first time a broadcast message from a neighbor node that 
belongs to another fragment, sets an internal flag,
and records the $id$ of the encountered fragment. 
It also records the identity of the node the message came from. 
If it receives additional broadcast messages from other fragments,
 it compares the $id$ delivered in the message to its registered one, 
according to definition~\ref{df:edge1}, and records in its internal 
variables the maximal size fragment it has encountered, and the identity 
of the node which sent the message. 

\item {\em Initiating FEEDBACK:} An {\em edge} node with no child nodes 
which has received a broadcast message from all of its neighbors, 
initiates a feedback message to its {\em parent} node. 
In the message it registers that it has encountered other fragments, 
and the maximal fragment $id$ it has encountered. 
The accumulated value of nodes under it is set to 1 and is sent. 
The node registers the number of neighboring nodes that belong 
to the  maximal fragment and reinitializes the variables that 
belong to the fragment-PIF.

\item {\em FEEDBACK process in a Fragment:} 
A node that receives a FEEDBACK message from one 
of its child nodes, compares the $id$ of the maximal neighboring 
fragment its child node knows of to the one it knows of. 
If the node does not know yet of another fragment encountered, 
or if the $id$ delivered by the child node is bigger than the current known 
neighboring fragment's $id$, according to definition~\ref{df:edge1}, 
the node registers the $id$ of the maximal neighbor fragment in its 
internal variables, and registers the child node's identity for a 
possible return path. It also adds the accumulated size of the 
sub-fragment under its child node to the current known count. 
A node that has received broadcast  messages from all of its 
neighboring nodes and feedback messages from all of its child nodes, 
and is not the {\em candidate} node, sends a feedback 
message to its {\em parent} node, which includes the accumulated size 
of the sub-fragment under it, and the maximal known neighbor fragment $id$, 
if any. The node then reinitializes the fragment-PIF variables. 
A {\em candidate} node that has received broadcast messages from 
all of its neighbor nodes and FEEDBACK messages from all of its child nodes, 
compares the maximal known neighbor {\em size}, as delivered, 
to its new counted size and then follows according to conditions 1 or 2. 
  \begin{enumerate}
  \item If its new size is at least twice the size of the maximal neighbor,
    then the candidate's fragment remains {\em active}. 
    The candidate then issues a new INFO message, 
    with the fragment's accumulated size, and sends its old 
    fragment size for recognition purposes. 
    It then updates its inner size variable to the sent one.
  \item Else, the candidate decides to become {\em inactive}, 
    and to join the maximal neighbor fragment. 
    It does so by sending the ACTION message along the node 
    path it has registered.
  \end{enumerate}
\item {\em ACTION Message Process:} A node which receives an ACTION message 
sends it along the path to the {\em edge} node. 
An {\em edge} node that receives an ACTION message from its {\em parent} 
node, does the following: It records the identity of the new fragment 
to join, and denotes the identity of the node from which it first heard 
of the other fragment as its {\em parent} node. 
It also initiates an INFO message, which contains the joined 
fragment's identity as the new identity, and the old fragment's identity, 
as sent in the ACTION message, as the old identity for recognition. 

\item {\em INFO Message Process:} A node that receives an INFO message, 
can receive it from the following sources: 
(a) From its parent in its fragment. In this case, it records the new 
fragment's {\em identity} or {\em size}, its parent node, 
and broadcasts the message. 
(b) From a neighbor in its fragment, in which case it registers the 
fact and ignores the message. (c) From an {\em edge} node in a joined 
fragment. If the node registers it as its parent node, it will reset 
the bit that indicates it has heard from it before, and await a FEEDBACK 
message from it. If not, it just ignores it.
\end{description}
\vspace{2mm}
\textbf{Properties of the Distributed Algorithm}\\
We introduce here a mapping of the states and transitions defined for the 
high level algorithm, as shown in~\figref{fig:FSM}, to the distributed 
algorithm described above, and establish the delays within the states:  
The delay {\em countdelay} within a fragment starts when the last 
{\em edge} node to send a feedback message has done so. It ends when the 
{\em candidate} node of the fragment receives all of the INFO messages 
from its neighbor nodes, and all of the FEEDBACK messages from its 
child nodes. \\
The delay {\em innerdelay} within a fragment is defined as the time 
passed from the initiation of an INFO message by the {\em source} node 
until all the nodes in its fragment receive the message. \\
For a fragment of size $k$, both {\em countdelay} and {\em innerdelay} are 
bounded by $k$ time units, as can be seen from Theorem~\ref{th:Pprop}. 
\lbl{pr:LDpitime}\\
The delay {\em innerdelay} is also the maximal time it takes the 
{\em candidate} node to send an ACTION message to the new {\em source} node. 
Again, this can be seen from Theorem~\ref{th:Pprop}. \\
The delay {\em workdelay} for a fragment starts when the last 
{\em edge} node to send a FEEDBACK message has done so, and ends when the 
last node in the fragment receives the new INFO message. \lbl{df:LDWD} \\
The delay {\em waitdelay} for a fragment starts when the last node within 
a fragment receives the INFO message, and ends when the last {\em edge} 
node sends the FEEDBACK message. \lbl{df:LDWTD} \\
{\em Cwork} is {\em true} if the last {\em edge} node within the fragment 
is able to send a FEEDBACK message. \\
{\em Cleader} is {\em true} if a candidate receives FEEDBACK messages from 
all of its fragment indicating that no other fragment was encountered. \\
{\em Ccease} is {\em true} if a candidate discovers it should become 
{\em inactive} and join another fragment. \\
{\em Cwait} is {\em true} if all the nodes within a fragment received 
a new INFO message originated at the {\em source} node.

\end{document}